\documentclass{article}
\usepackage{spconf,amsmath}

\usepackage{multirow}
\usepackage{graphicx}
\usepackage{amsmath}
\usepackage[psamsfonts]{amssymb}
\usepackage{amsxtra}
\usepackage{threeparttable}
\usepackage{amsfonts}
\usepackage{mathrsfs}
\usepackage{arydshln}
\usepackage{cite}
\usepackage{enumitem}
\usepackage{url}

\makeatletter
\let\MYcaption\@makecaption
\makeatother
\usepackage{subcaption}
\makeatletter
\let\@makecaption\MYcaption
\makeatother
\makeatletter
\renewcommand\section{\@startsection{section}{1}{\z@}
                      {0.5ex \@plus 0ex \@minus -2ex}
                      {0.5ex \@plus 0ex}
                      {\normalfont\Large\bfseries}}
\renewcommand\subsection{\@startsection{subsection}{2}{\z@}
                      {0.5ex \@plus 0ex \@minus -2ex}
                      {0.5ex \@plus 0ex}
                      {\normalfont\large\bfseries}}
\renewcommand\subsubsection{\@startsection{subsubsection}{3}{\z@}
                      {0.5ex \@plus 0ex \@minus -2ex}
                      {0.5ex \@plus 0ex}
                      {\normalfont\normalsize\bfseries}}
\def\@listi{\leftmargin\leftmargini
            \parsep 1.0pt
            \topsep 0.2\baselineskip \@minus 0.1\baselineskip
            \itemsep 1.0pt \relax}
\let\@listI\@listi
\makeatother

\newcommand{\myvector}[1]{\boldsymbol{#1}}
\newcommand{\mymatrix}[1]{\mathrm{#1}}
\newcommand{\mytensor}[1]{\boldsymbol{\mathrm{#1}}}
\newcommand{\myset}[1]{\mathbb{#1}}

\newcounter{num}

\title{Fixed smooth convolutional layer for avoiding checkerboard artifacts in CNNs}
\name{Yuma Kinoshita and Hitoshi Kiya}
\address{Tokyo Metropolitan University, Tokyo, Japan}
\begin{document}\sloppy
\setlength{\parskip}{0.0pt}
\setlength{\tabcolsep}{1.0pt}
\setlength{\textfloatsep}{0.0pt}
\setlength{\floatsep}{1.0pt}
\setlength{\intextsep}{0.0pt}
\setlength{\abovecaptionskip}{0.0pt}
\setlength{\belowcaptionskip}{2.0pt}
\setlength{\dblfloatsep}{0.0pt}
\setlength{\dbltextfloatsep}{1.0pt}
\setlength{\lineskiplimit}{0.0pt}
\setlength{\lineskip}{0.0pt}
\setlength{\abovedisplayskip}{0.0pt}
\setlength{\belowdisplayskip}{0.0pt}
\setlength{\abovedisplayshortskip}{0.0pt}
\setlength{\belowdisplayshortskip}{0.0pt}
\ninept
\maketitle
\begin{abstract}
  In this paper, we propose a fixed convolutional layer
  with an order of smoothness not only for avoiding
  checkerboard artifacts in convolutional neural networks (CNNs)
  but also for enhancing the performance of CNNs,
  where the smoothness of its filter kernel can be controlled
  by a parameter.
  It is well-known that a number of CNNs generate checkerboard artifacts
  in both of two process: forward-propagation of upsampling layers
  and backward-propagation of strided convolutional layers.
  The proposed layer can perfectly prevent checkerboard artifacts
  caused by strided convolutional layers or upsampling layers
  including transposed convolutional layers.
  In an image-classification experiment with four CNNs:
  a simple CNN, VGG8, ResNet-18, and ResNet-101,
  applying the fixed layers to these CNNs is shown to
  improve the classification performance of all CNNs.
  In addition, the fixed layer are applied to
  generative adversarial networks (GANs), for the first time.
  From image-generation results,
  a smoother fixed convolutional layer is demonstrated to enable us
  to improve the quality of images generated with GANs.
\end{abstract}
\begin{keywords}
  Checkerboard artifacts, Convolutional neural networks, Deep learning,
  Fixed convolutional layers, Generative adversarial networks
\end{keywords}
\renewcommand{\thefootnote}{\fnsymbol{footnote}}
\footnote[0]{This work was supported by JSPS KAKENHI Grant Number JP18J20326.}
\renewcommand{\thefootnote}{\arabic{footnote}}

\section{Introduction}
  Convolutional neural networks (CNNs)~\cite{lecun1989generalization, krizhevsky2012imagenet},
  which are a kind of deep neural networks (DNNs),
  have attracted attention because of their outstanding performance
  and have widely been used in many fields: image processing, natural language processing,
  acoustic/speech processing, and more.
  In research fields of image-to-image translation problems,
  e.g., image super-resolution,
  it is well-known that forward-propagation of upsampling layers
  including transposed convolutional layers
  causes images to be distorted by checkerboard artifacts~\cite{odena2016deconvolution}.
  In addition, checkerboard artifacts are also caused
  by backward-propagation of downsampling layers including strided convolutional layers
  ~\cite{odena2016deconvolution}.
  CNN architectures usually have upsampling layers and/or have downsampling layers,
  such as VGG~\cite{simonyan2014very}, ResNet~\cite{he2016deep},
  and U-Net~\cite{ronneberger2015unet},
  for increasing and/or reducing the spatial sampling rate of feature maps,
  respectively~\cite{goodfellow2016deep}.
  For this reason,
  checkerboard artifacts affect most commonly-used CNNs.

  To overcome checkerboard artifacts caused by upsampling layers,
  a few research has been done to reduce the effects of checkerboard artifacts
  ~\cite{johnson2016perceptual, aitken2017checkerboard,
  odena2016deconvolution, gao2017pixel, wojna2017devil, sugawara2018super, sugawara2019checkerboard}.
  In particular, 
  Sugawara et al.~\cite{sugawara2018super, sugawara2019checkerboard}
  gave us two approaches to perfectly prevent checkerboard artifacts
  by extending a condition for avoiding checkerboard artifacts
  in linear multirate systems
  ~\cite{harada1998multidimensional, tamura1998design, harada1998multidimensional_trans,
  iwai2010methods}.
  However, the literature~\cite{sugawara2018super, sugawara2019checkerboard}
  mainly focuses on checkerboard artifacts caused by upsampling layers,
  but there are few discussion about artifacts caused by downsampling layers.
  In addition, Sugawara et al. did not consider CNNs that have both
  upsampling and downsampling layers, such as generative adversarial networks (GANs)
  ~\cite{goodfellow2014generative}.
  Furthermore, in Sugawara's method,
  only a zero-order hold kernel that has the lowest smoothness
  is used for avoiding checkerboard artifacts.
  Note that the smoothness of filter kernels is different from
  what has widely been considered for GANs,
  i.e., Lipschitz constraints
  ~\cite{arjovsky2017wasserstein, gulrajani2017improved, miyato2018spectral}.
  Hence, the effects of the smoothness on the performance of CNNs
  has never been discussed so far.
  
  Because of such a situation,
  in this paper,
  we propose a novel fixed convolutional layer with an order of smoothness
  not only for perfectly avoiding checkerboard artifacts caused
  by upsampling and downsampling layers
  but also for enhancing the performance of CNNs.
  A filter kernel of the proposed layer is given by convolving a zero-order hold kernel
  multiple times.
  When using the proposed layer,
  we can control the smoothness of its filter kernel by changing
  the number of the convolution, referred to as \textit{the order of smoothness}.
  In addition, we apply the proposed fixed layer to GANs for the first time,

  We performed an image-classification experiment and an image-generation experiment
  in order to evaluate the effectiveness of the proposed fixed smooth convolutional layers.
  In the image-classification experiment with four CNNs:
  a simple CNN, VGG8, ResNet-18, and ResNet-101,
  applying the proposed layer is shown to improve the classification performance
  of all CNNs.
  From the image-generation results,
  a smoother fixed convolutional layer enables us not only
  to prevent checkerboard artifacts in GANs
  but also to improve the quality of images generated with GANs.

\section{Preparation}
  In this section, we briefly summarize CNNs and checkerboard artifacts,
  where we only focus on 2-D convolutional layers for simplicity.
  We use notations shown in Table \ref{tab:notation} throughout this paper.
  \begin{table}[!t]
    \centering
    \footnotesize
    \caption{Notation}
    \begin{tabular}{cp{0.8\columnwidth}} \hline\hline
      Symbol & Definition \\ \hline
      $a$ & A scalar \\
      $\myvector{a}$ & A vector \\
      $\myvector{a}_i$ & Element $i$ of vector $\myvector{a}$, with indexing starting at $1$ \\
      $\mymatrix{A}$ & A matrix \\
      $\mymatrix{A}_{i, j}$ & Element $i, j$ of matrix $\mymatrix{A}$ \\
      $\mymatrix{A}_{i, :}$ & Row $i$ of matrix $\mymatrix{A}$ \\
      $\mymatrix{A}_{:, i}$ & Column $i$ of matrix $\mymatrix{A}$ \\
      $\mytensor{A}$ & A 3-D or higher-dimensional tensor \\
      $\mytensor{A}_{i, j, k}$ & Element $i, j, k$ of 3-D tensor $\mytensor{A}$ \\
      $\mytensor{A}_{:, :, i}$ & 2-D slice of 3-D tensor $\mytensor{A}$ \\
      $\mytensor{V}$ & A 3-D tensor with a size of
        \textit{channel} $\times$ \textit{height} $\times$ \textit{width},
        which denotes an input feature map for a layer. \\
      $\mytensor{K}$ & A 4-D tensor with a size of
        \textit{output channel} $\times$ \textit{input channel}
        $\times$ \textit{height} $\times$ \textit{width},
        which denotes a filter kernel (weights) of a convolutional layer. \\
      $\myvector{b}$ & A vector with a size of \textit{output channel}
        which denotes a bias of a convolutional layer. \\
      $\mytensor{Z}$ & A 3-D tensor
        with a size of \textit{output channel} $\times$ \textit{height} $\times$ \textit{width},
        which denotes an output feature map of a layer. \\
      $\mymatrix{I}_i$ & A single-channel image which is a 2-D slice $\mytensor{V}_{i, :, :}$
        of $\mytensor{V}$. \\
      $\mymatrix{h}_{i, j}$ & A 2-D filter which is
        a 2-D slice $\mytensor{K}_{i, j, :, :}$ of filter kernel $\mytensor{K}$. \\
      $\mathcal{L}$ & A loss function \\
      $\myset{Z^+}$ & The set of non-negative integers \\
      $\mymatrix{A} * \mymatrix{B} $ & The convolution on two matrices
        $\mymatrix{A}$ and $\mymatrix{B}$ \\
      \hline
    \end{tabular}
    \label{tab:notation}
  \end{table}

\subsection{Convolutional neural networks}
\begin{figure}[!t]
  \centering
  \begin{subfigure}[t]{0.35\hsize}
    \centering
    \includegraphics[width=\columnwidth]{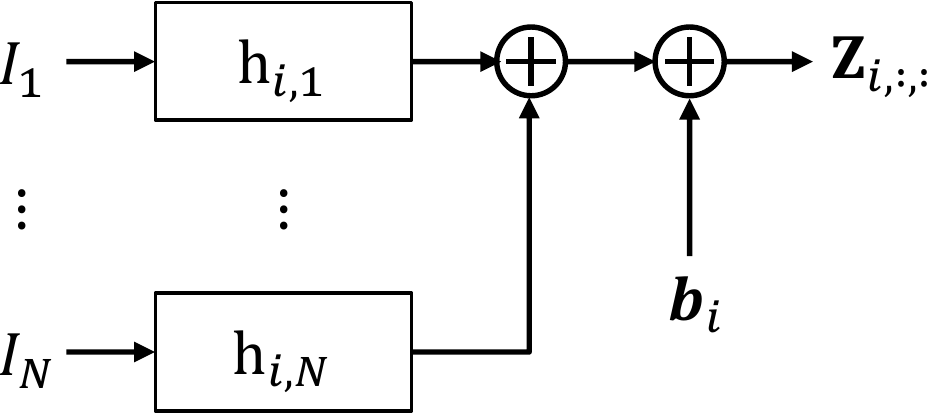}
    \caption{Convolution \label{fig:block_conv}}
  \end{subfigure}\\
  \begin{subfigure}[t]{0.45\hsize}
    \centering
    \includegraphics[width=\columnwidth]{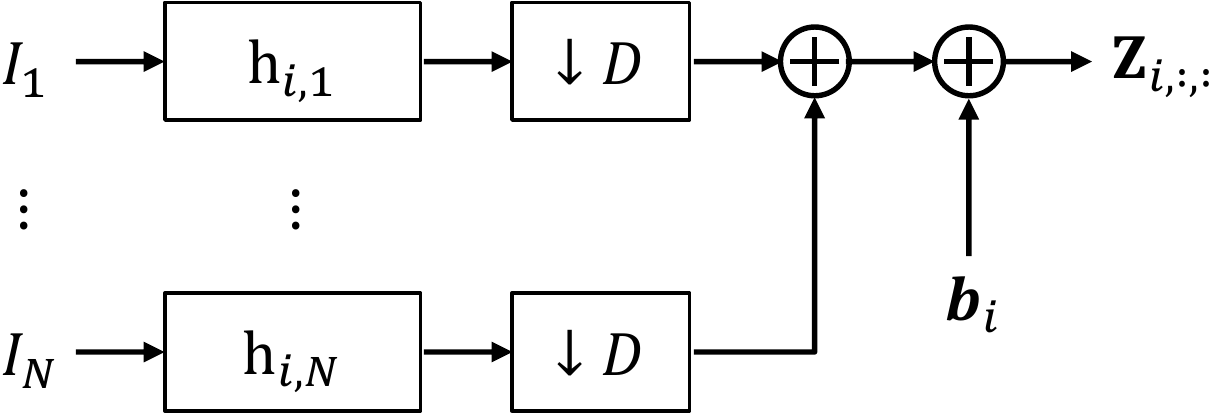}
    \caption{Strided convolution \label{fig:block_strided}}
  \end{subfigure}
  \begin{subfigure}[t]{0.45\hsize}
    \centering
    \includegraphics[width=\columnwidth]{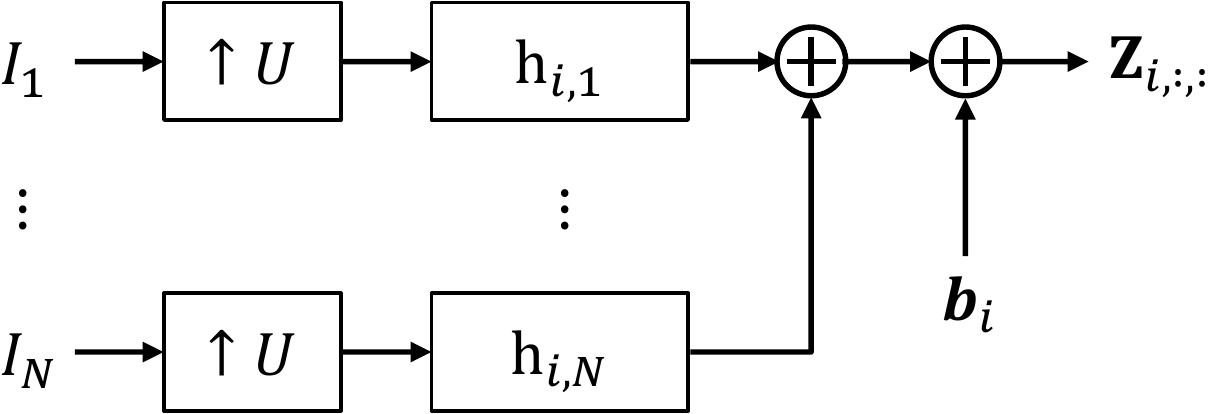}
    \caption{Transposed convolution \label{fig:block_trans}}
  \end{subfigure}
  \vspace{-2ex}
  \caption{Block diagram of convolutional layers \label{fig:block}}
\end{figure}

  A CNN is one of DNNs that has one or more convolutional layer(s).
  Calculation in a convolutional layer is given as
  \begin{equation}
    \mytensor{Z}_{i, j, k} = \sum_{l, m, n} \mytensor{V}_{l, j+m-1, k+n-1}
                                            \mytensor{K}_{i, l, m, n}
                           + \myvector{b}_i,
    \label{eq:conv_layer}
  \end{equation}
  where $\mytensor{V}$ denotes an input feature map,
  $\mytensor{K}$ is a filter kernel,
  and $\myvector{b}$ is a bias.
  Focusing on the $i$-th channel of an output feature map,
  the calculation of eq. (\ref{eq:conv_layer}) can be illustrated
  as a block-diagram by using a 2-D filter $\mymatrix{h}_{i, j} = \mytensor{K}_{i, j, :, :}$
  and a single-channel image $\mymatrix{I}_j$ = $\mytensor{V}_{j, :, :}$,
  as shown in Fig. \ref{fig:block}(\subref{fig:block_conv}).

  CNNs usually include upsampling layers and/or downsampling layers,
  to increase and/or reduce
  the spatial sampling rate (or the resolution) of feature maps, respectively.
  Typical downsampling layers are: strided convolutional layers,
  average pooling layers, and max pooing layers~\cite{goodfellow2016deep}.
  Here, average pooling layers and max pooling layers can be replaced
  with strided convolutional layers
  because:
  \begin{itemize}[nosep]
    \item An average pooling layer is a particular kind of a strided convolutional layer.
    \item Replacing a max-pooling layer has little effect on the performance of CNNs
  ~\cite{springenberg2014striving}.
  \end{itemize}
  For these reasons, we focus on strided convolutional layers
  as downsampling layers.
  A strided convolutional layer is defined by the following equation:
  \begin{align}
    \mytensor{Z}_{i, j, k} &= c(\mytensor{K}, \mytensor{V}, s)_{i, j, k}
                            + \myvector{b}_i \nonumber \\
                           &= \sum_{l, m, n} \mytensor{V}_{l, (j-1)s+m, (k-1)s+n}
                                             \mytensor{K}_{i, l, m, n}
                            + \myvector{b}_i,
    \label{eq:strided_conv_layer}
  \end{align}
  where $s \in \myset{Z^+}$ is a parameter called \textit{stride}.
  Focusing on the $i$-th channel of an output feature map,
  the calculation of a strided convolutional layer corresponds to
  downsampling signals with a downsampling rate $D=s$ after
  convolving an input feature map $\mymatrix{I}_j$ with a filter $\mymatrix{h}_{i, j}$
  [see Fig. \ref{fig:block}(\subref{fig:block_strided})].

  Upsampling layers includes a transposed convolutional layer,
  a sub-sampling layer~\cite{shi2016real},
  and a resize convolution layer~\cite{odena2016deconvolution}.
  In these layers,
  the most commonly-used transposed convolutional layer is given as
  \begin{align}
    \mytensor{Z}_{i, j, k} &= t(\mytensor{K}, \mytensor{V}, s)_{i, j, k}
                            + \myvector{b}_i \nonumber \\
                           &= \sum_{\substack{l, m \\ \mathrm{s.t.} \\ (l-1)s+m=j}}
                              \sum_{\substack{n, p \\ \mathrm{s.t.} \\ (n-1)s+p=k}}
                              \sum_q \mytensor{V}_{q, l, n}
                                     \mytensor{K}_{q, i, m, p}
                            + \myvector{b}_i.
    \label{eq:trans_conv_layer}
  \end{align}
  In contrast to a strided convolutional layer,
  the calculation for obtaining the $i$-th channel of an output feature map
  of a transposed convolutional layer corresponds to
  upsampling an input feature map $\mymatrix{I}_j$ with an upsampling rate $U=s$
  and then convolving the resulting feature map with a filter $\mymatrix{h}_{i, j}$
  [see Fig. \ref{fig:block}(\subref{fig:block_trans})].

\subsection{Upsampling and checkerboard artifacts\label{sec:checkerboard}}
\begin{figure}[!t]
  \centering
  \begin{subfigure}[t]{0.45\hsize}
    \centering
    \includegraphics[width=\columnwidth]{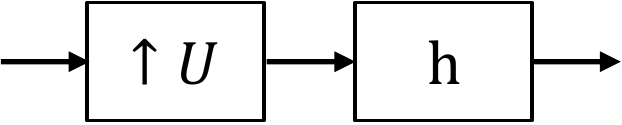}
    \caption{Interpolator \label{fig:lin_interpolator}}
  \end{subfigure}
  \begin{subfigure}[t]{0.45\hsize}
    \centering
    \includegraphics[width=\columnwidth]{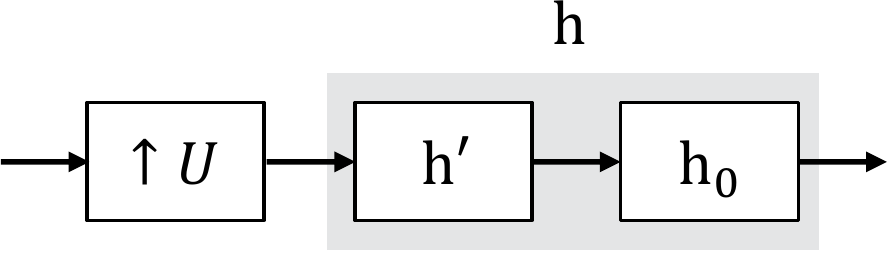}
    \caption{w/o checkerboard artifact \label{fig:lin_wo_checkerboard}}
  \end{subfigure}
  \vspace{-2ex}
  \caption{Linear interpolator and avoiding checkerboard artifacts \label{fig:linear}}
\end{figure}

  Checkerboard artifacts have been studied as a distortion
  caused by using upsamplers in linear multi-rate systems
  ~\cite{harada1998multidimensional, tamura1998design, harada1998multidimensional_trans,
  iwai2010methods}.
  Figure \ref{fig:linear}(\subref{fig:lin_interpolator}) shows
  a simple linear multi-rate system called \textit{linear interpolator}
  that consists of an upsampler $\uparrow U$ with an upsampling rate $U$
  and a linear time-invariant filter $\mymatrix{h}$.
  To avoid checkerboard artifacts caused by the linear interpolator,
  it is necessary and sufficient that its filter $\mymatrix{h}$
  satisfies the following condition:
  \begin{equation}
    \mymatrix{h} = \mymatrix{h}_0 * \mymatrix{h}',
    \label{eq:general_condition}
  \end{equation}
  where $\mymatrix{h}_0$ and $\mymatrix{h}'$ are the zero-order hold kernel
  with a size of $U \times U$
  and another linear time-invariant filter, respectively,
  and $*$ means the convolution on two matrices.

\begin{figure}[!t]
  \centering
  \begin{subfigure}[t]{0.40\hsize}
    \centering
    \includegraphics[width=\columnwidth]{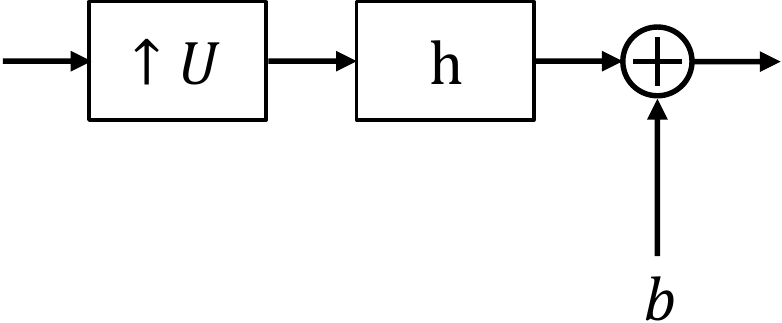}
    \caption{Interpolator w/ bias\label{fig:nonlin_interpolator}}
  \end{subfigure}\\
  \begin{subfigure}[t]{0.40\hsize}
    \centering
    \includegraphics[width=\columnwidth]{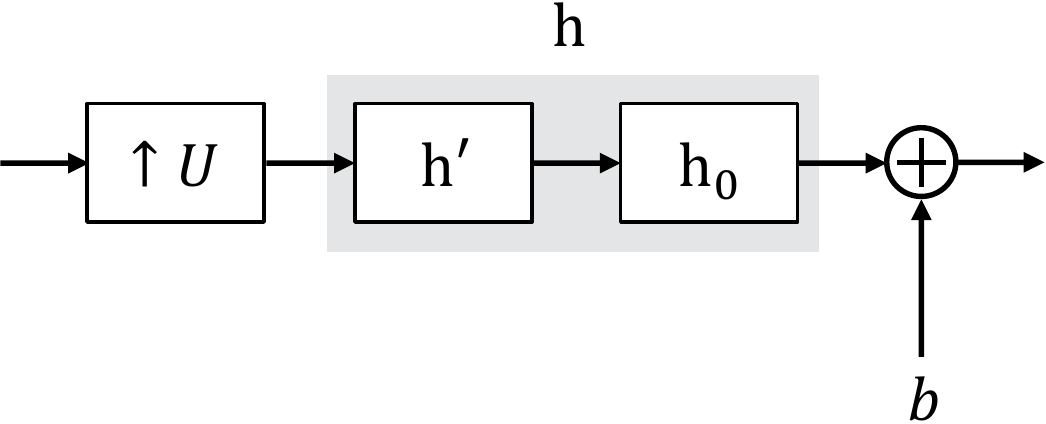}
    \caption{w/o checkerboard artifact (Approach 1) \label{fig:nonlin_wo_checkerboard_1}}
  \end{subfigure}
  \begin{subfigure}[t]{0.40\hsize}
    \centering
    \includegraphics[width=\columnwidth]{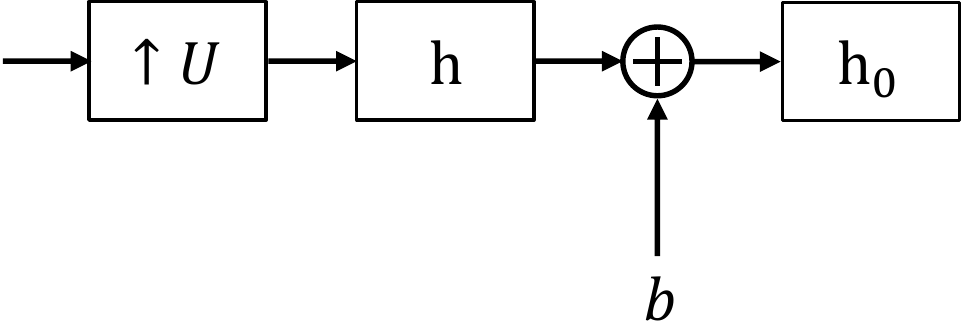}
    \caption{w/o checkerboard artifact (Approach 2) \label{fig:nonlin_wo_checkerboard_2}}
  \end{subfigure}
  \vspace{-2ex}
  \caption{Non-linear interpolator with bias
    and avoiding checkerboard artifacts \label{fig:nonlinear}}
\end{figure}

  A transposed convolutional layer has non-linear interpolators having a bias $b$,
  as in Fig. \ref{fig:nonlinear}(\subref{fig:nonlin_interpolator}).
  Sugawara et al.~\cite{sugawara2018super, sugawara2019checkerboard}
  realized the perfect avoidance of checkerboard artifacts
  in non-linear interpolators by using the following two approaches:
  \begin{enumerate}[nosep, label=\textbf{Approach \arabic*}, leftmargin=*]
    \item Insert the zero-order hold kernel $\mymatrix{h}_0$ after
      a filter $\mymatrix{h}'$
      [see Fig. \ref{fig:nonlinear}(\subref{fig:nonlin_wo_checkerboard_1})].
      This is equivalent to applying the condition in eq. (\ref{eq:general_condition})
      to $\mymatrix{h}$.
    \item Insert $\mymatrix{h}_0$ after adding a bias $\myvector{b}$,
      without any constraints on a filter $\mymatrix{h}$
      [Fig. \ref{fig:nonlinear}(\subref{fig:nonlin_wo_checkerboard_2})].
  \end{enumerate}
  Note that Approaches 1 and 2 cannot be equivalently converted to each other.

  However, the literature~\cite{sugawara2019checkerboard}
  mainly focuses on checkerboard artifacts caused by upsampling layers,
  i.e., artifacts in the forward-propagation,
  but there are few discussion about artifacts caused by downsampling layers,
  i.e., artifacts in the backward-propagation.
  In addition, Sugawara et al. did not consider CNNs
  that have both upsampling and downsampling layers, such as GANs.
  Furthermore, in Sugawara's method,
  a zero-order hold kernel that has the lowest smoothness
  is used for avoiding checkerboard artifacts,
  although we have a choice of the smoothness of filter kernels.

\section{Proposed method}
\begin{figure}[!t]
  \centering
  \begin{subfigure}[t]{0.95\hsize}
    \centering
    \includegraphics[width=\columnwidth]{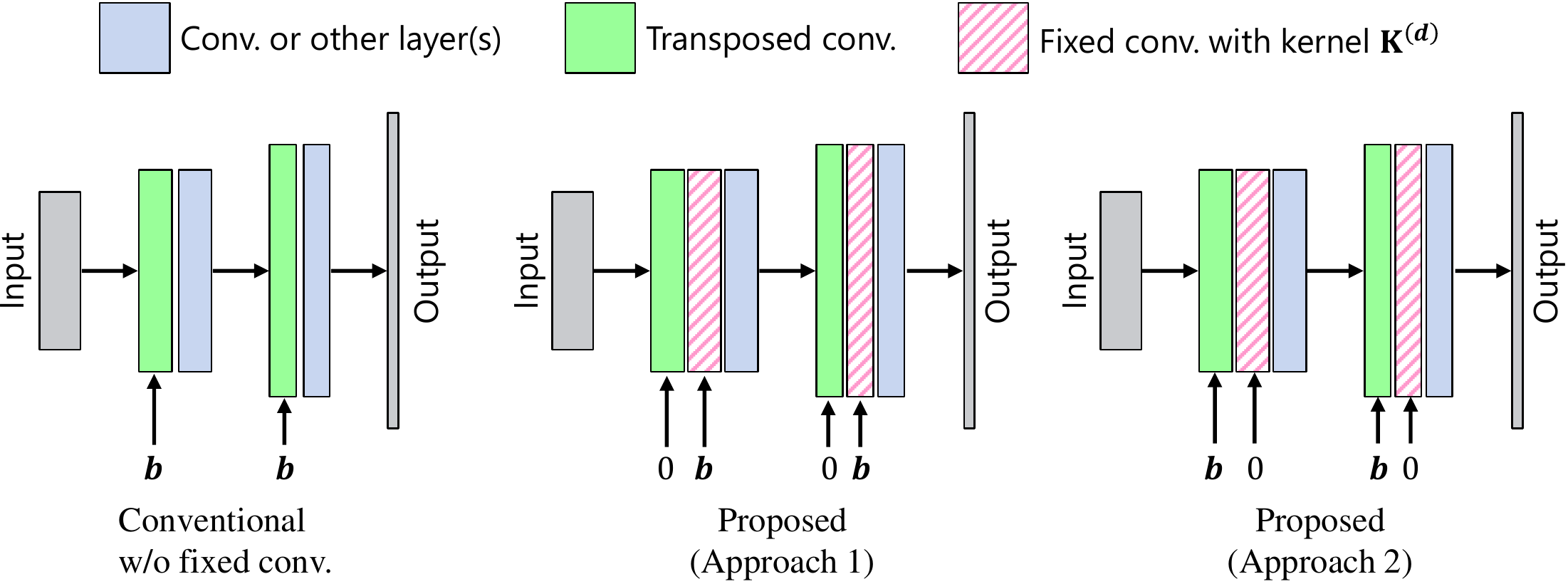}
    \caption{For transposed convolution \label{fig:proposed_up}}
  \end{subfigure}\\
  \begin{subfigure}[t]{0.95\hsize}
    \centering
    \includegraphics[width=\columnwidth]{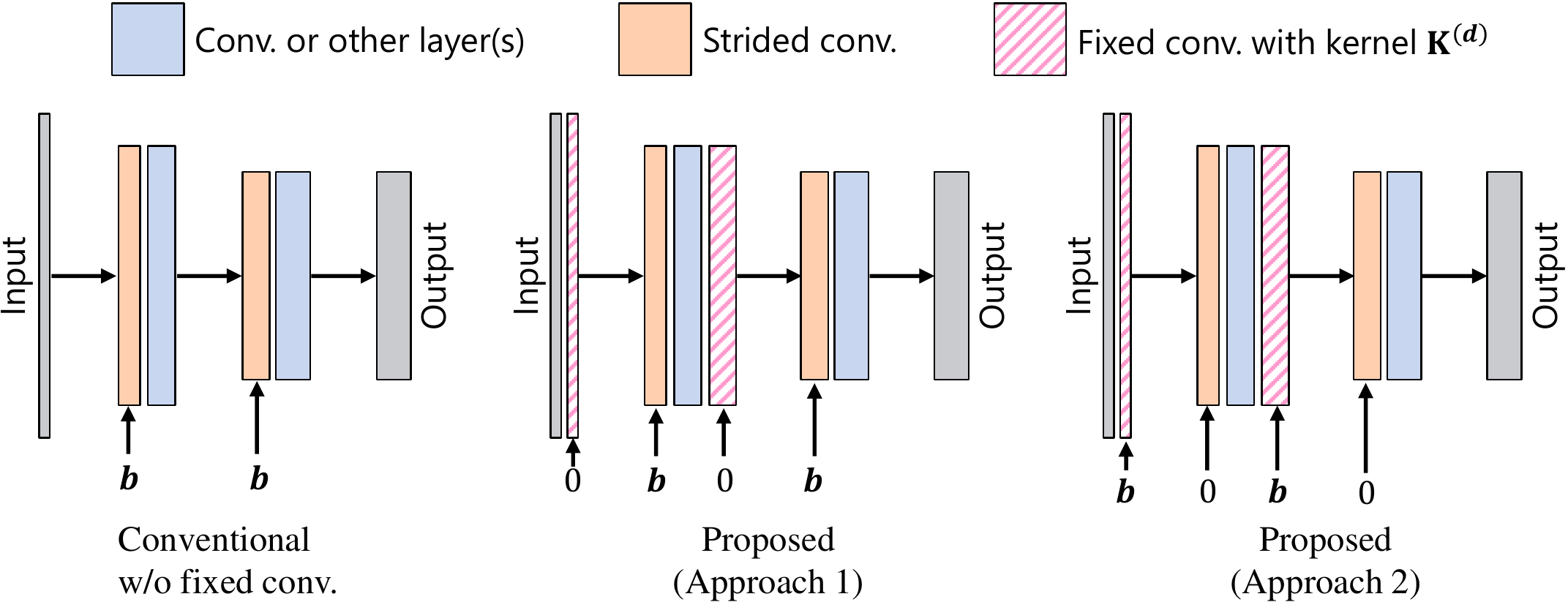}
    \caption{For strided convolution \label{fig:proposed_down}}
  \end{subfigure}
  \vspace{-2ex}
  \caption{Usage of proposed fixed convolution layer \label{fig:proposed}}
\end{figure}

  Figure \ref{fig:proposed} shows an usage of the fixed smooth convolutional layer
  that we propose.
  The proposed layer is a (depth-wise) convolutional layer
  and the smoothness of its filter kernel $\mytensor{K}^{(d)}$
  is controlled by a parameter $d$ called \textit{the order of smoothness}.
  Similarly to Sugawara's method in Section \ref{sec:checkerboard},
  there are two approaches to apply the proposed layer
  to a transposed convolutional layer or a strided convolutional layer,
  depending on the location of a bias $\myvector{b}$.

\subsection{Checkerboard artifacts in CNNs}
  Checkerboard artifacts in CNNs can be classified into
  artifacts in the forward-propagation and ones in the backward-propagation.

  Checkerboard artifacts in the forward-propagation are caused by
  interpolators in transposed convolutional layers
  as shown in Fig. \ref{fig:block}(\subref{fig:block_trans}),
  where the forward-propagation means
  a process calculating a new feature map $\mytensor{Z}$
  from an input feature map $\mytensor{V}$
  and passing the new feature map to the next layer.
  All filter kernels in transposed convolutional layers
  have to be trained to satisfy the condition in eq. (\ref{eq:general_condition}),
  to prevent checkerboard artifacts.
  However, checkerboard artifacts cannot be prevented in practice
  because it is difficult to perfectly satisfy this condition
  by using an optimization algorithm.
  To resolve the issue, the use of fixed layers is required.
  Conventional CNNs with upsampling layers,
  which are mainly used for image super-resolution,
  distort an output feature map of each layer and an output image
  due to the artifacts.
  In addition, the artifacts affect training CNN models
  because the distorted output images are used for evaluating a loss function,
  for training the CNNs.

  Checkerboard artifacts are also caused in the backward-propagation
  of a strided convolutional layer.
  The backward-propagation is
  a process calculating gradients of a loss function $\mathcal{L}$
  with respect to parameters in a CNN, in order from the output layer
  to the input layer.
  In a strided convolutional layer,
  calculation of the gradient with respect to an input feature map $\mytensor{V}$
  is done by using transposed convolution $t(\cdot)$ as
  \begin{equation}
    \frac{\partial}{\partial \mytensor{V}_{i, j, k}} \mathcal{L}(\mytensor{V}, \mytensor{K})
      = t(\mytensor{K}, \mytensor{G}, s)_{i, j, k},
    \label{eq:grad_strided}
  \end{equation}
  where a 3-D tensor $\mytensor{G}$ of a gradient is given by an output-side layer
  and $\mytensor{K}$ denotes a filter kernel that the strided convolutional layer has.
  For this reason,
  checkerboard artifacts appear in calculated gradients
  of CNNs with downsampling layers, which are generally used for image classification.
  Therefore,
  a filter kernels $\mytensor{K}$ and a bias $\myvector{b}$ in
  these CNNs will be affected by the checkerboard artifacts in gradients.

  In addition, there are CNNs including both upsampling and downsampling layers,
  e.g., GANs~\cite{goodfellow2014generative} and the U-Net~\cite{ronneberger2015unet},
  which are generally used for image generation and image-to-image translation.
  These CNNs are affected by the checkerboard artifacts mentioned above.

\subsection{Fixed smooth convolutional layer}
  A filter kernel $\mytensor{K}^{(d)}$ of the proposed layer is obtained
  by convolving $\mymatrix{h}_0$ multiple times, as
  \begin{equation}
    \mytensor{K}^{(d)}_{i, i, :, :} =
    \begin{cases}
      \mymatrix{h}_0 & (d = 0 ) \\
      \mytensor{K}^{(d-1)}_{i, i, :, :} * \mymatrix{h}_0 & (d > 1)
    \end{cases},
  \end{equation}
  where a parameter $d$, referred to as the order of smoothness,
  controls the smoothness of $\mytensor{K}^{(d)}$.
  When the proposed layer is applied to an upsampling layer with an upsampling rate $U$,
  the kernel size for $\mymatrix{h}_0$ is given as $U \times U$.
  In contrast,
  the kernel size is given as $D \times D$
  when the proposed layer is applied to a downsampling layer with a downsampling rate $D$.
  By using a filter kernel $\mytensor{K}^{(d)}$ and a trainable bias $\myvector{b}$,
  an output feature map $\mytensor{Z}$ of the proposed layer
  can be written as
  \begin{equation}
    \mytensor{Z}_{i, j, k} = \sum_{m, n} \mytensor{V}_{i, j+m-1, k+n-1}
                                         \mytensor{K}^{(d)}_{i, i, m, n} + \myvector{b}_i.
  \end{equation}

\subsection{Avoiding checkerboard artifacts by the proposed layer}
  Checkerboard artifacts caused by a transposed convolutional layer
  can perfectly be prevented by using the proposed layer as follows
  [see also Fig. \ref{fig:proposed}(\subref{fig:proposed_up})]:
  \begin{enumerate}[nosep, label=\textbf{Approach \arabic*}, leftmargin=*]
    \item Fix a bias in a transposed convolutional layer as $0$,
      and insert the proposed layer having a trainable bias $\myvector{b}$ after
      the transposed convolutional layer.
    \item Insert the proposed layer having a bias which is fixed as 0
      after a transposed convolutional layer having a trainable bias $\myvector{b}$.
  \end{enumerate}
  These approaches correspond to Approaches 1 and 2 in Section \ref{sec:checkerboard},
  respectively.
  The use of Approach 1 allows us to reduce computational costs
  since the redundancy of interpolators can be removed by the polyphase decomposition,
  but it will negate the effect of the bias $\myvector{b}$
  since the batch normalization (BN) is generally applied
  after each transposed convolutional layer.
  In contrast,
  under the use of Approach 2,
  the bias $\myvector{b}$ is not affected by the BN
  but its computational costs cannot be reduced.

  Checkerboard artifacts caused by a strided convolutional layer can also be prevented
  by inserting the proposed layer before the strided convolutional layer,
  in an opposite way to a transposed convolutional layer,
  because checkerboard artifacts in the strided convolutional layer
  occur in the backward-propagation
  [see \ref{fig:proposed}(\subref{fig:proposed_down})].
  Similarly to the case of transposed convolutional layers,
  we can consider two approaches in terms of the location of a bias.

  Both Approaches 1 and 2 for the proposed fixed layer
  can prevent checkerboard artifacts caused
  by both transposed convolutional layers and strided convolutional layers,
  under any order of smoothness $d$.
  In the next section,
  we will confirm that both Approaches 1 and 2 enhance the performance of CNNs
  and evaluate the quality of images generated by GANs
  under the use of the proposed layer having a high-order of smoothness.

\section{Simulation}
  To evaluate the effectiveness of avoiding checkerboard artifacts by the proposed layer,
  we performed two simulations: image classification and image generation.

\subsection{Image classification}
  We evaluated the classification accuracy of four CNNs with/without
  the proposed layer,
  in order to confirm the effectiveness of avoiding checkerboard artifacts in
  the backward-propagation.
  Here, we set all filter kernels of the proposed fixed layers as $\mytensor{K}^{(0)}$.
  The following are the four CNNs used in this simulation:
  a simple CNN illustrated in Table \ref{tab:simple_cnn}, VGG8, ResNet-18, and ResNet-101,
  where ReLU in the table denotes
  the rectified linear unit activation function~\cite{glorot2011deep}.
  \begin{table}[!t]
    \centering
    \footnotesize
    \caption{Architecture of simple CNN.
      \textit{ch} denotes number of output channels.}
    \begin{tabular}{cccc} \hline\hline
      Layer & Stride $s$ & Kernel size & \textit{ch} \\ \hline
      Conv. + BN + ReLU & 2 & $3 \times 3$ & 64 \\
      Conv. + BN + ReLU & 2 & $3 \times 3$ & 128 \\
      Conv. + BN + ReLU & 2 & $3 \times 3$ & 256 \\
      Conv. + BN + ReLU & 2 & $3 \times 3$ & 512 \\
      Conv. + BN + ReLU & 1 & $2 \times 2$ & 10 \\
      \hline
    \end{tabular}
    \label{tab:simple_cnn}
  \end{table}
  In this simulation,
  we used the CIFER10 dataset~\cite{krizhevsky2009learning}
  without any data augmentation
  for training and testing CNNs.
  Each CNN was trained with 300 epochs by using the Adam optimizer~\cite{kingma2014adam},
  where parameters in Adam were set as an initial learning rate of 0.1,
  $\beta_1=0.9$, and $\beta_2=0.999$,
  and we utilized an initial learning rate of 0.01 only for VGG8.
  In addition, each learning rate was multiplied by $1/10$
  when the number of epoch reached 150 and 225.

\begin{figure}[!t]
  \centering
  \begin{subfigure}[t]{0.35\hsize}
    \centering
    \includegraphics[width=\columnwidth]{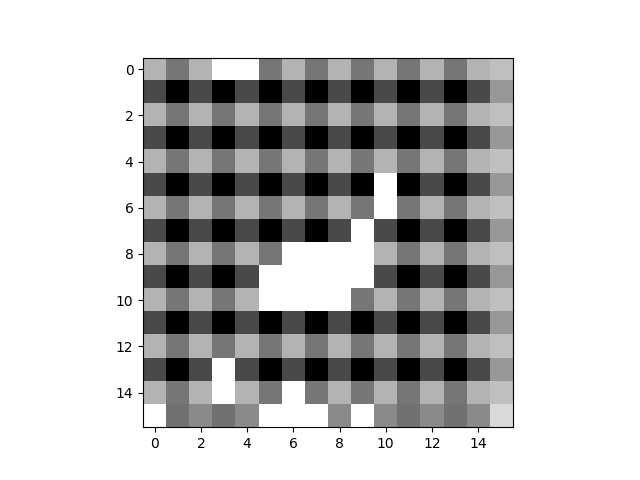}
    \caption{w/o fixed conv. layer \label{fig:step_response_conv}}
  \end{subfigure}
  \begin{subfigure}[t]{0.35\hsize}
    \centering
    \includegraphics[width=\columnwidth]{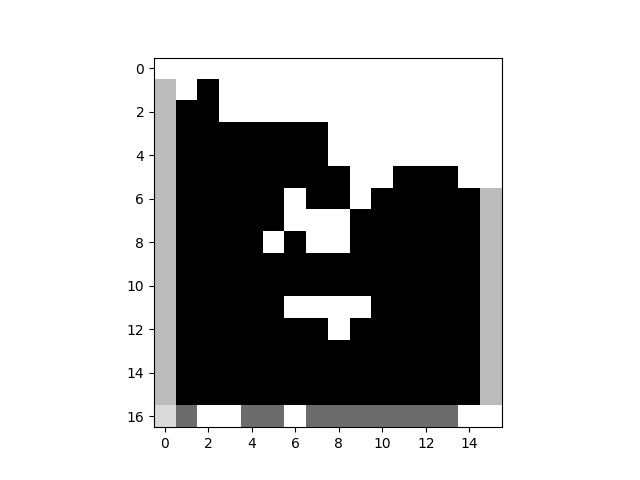}
    \caption{with fixed conv. layer \label{fig:step_response_prop}}
  \end{subfigure}
  \vspace{-2ex}
  \caption{Unit step response of backward-propagation of
    second convolutional layer in simple CNN \label{fig:step_response}}
\end{figure}

  Figure \ref{fig:step_response} shows
  the unit-step response of the backward-propagation of
  the second convolutional layer in the simple CNN with/without the proposed layers.
  From Fig. \ref{fig:step_response}(\subref{fig:step_response_conv}),
  we can confirm that 
  checkerboard artifacts appeared in calculated gradients
  when the simple CNN without the proposed layer was used.
  The artifacts were perfectly prevented by the proposed layers
  [see \ref{fig:step_response}(\subref{fig:step_response_conv})].

  Table \ref{tab:classification_results} illustrates
  the classification accuracy of the four CNNs.
  From Table \ref{tab:classification_results},
  the proposed fixed layer improved the accuracy for the test set of all four CNNs.
  This improvement was confirmed in both Approaches 1 and 2
  for avoiding checkerboard artifacts.
  The results illustrate that conventional CNNs for image classification
  were affected by checkerboard artifacts,
  and preventing the artifacts is effective for improving their performance.
  \begin{table}[!t]
    \centering
    \footnotesize
    \caption{Image-classification accuracy for CIFER10 (train/test)}
    \begin{tabular}{l|ccc}\hline\hline
      Network & Conv. & Prop. (Approach 1) & Prop. (Approach 2) \\ \hline
      Simple CNN & 0.999/0.585 & 0.999/0.645 & 0.976/\textbf{0.671} \\
      VGG8 & 1.000/0.795 & 1.000/\textbf{0.840} & 1.000/0.839 \\
      ResNet-18 & 1.000/0.812 & 1.000/\textbf{0.862} & 1.000/\textbf{0.862} \\
      ResNet-101 & 1.000/0.852 & 1.000/\textbf{0.863} & 1.000/\textbf{0.863} \\
      \hline
    \end{tabular}
    \label{tab:classification_results}
  \end{table}

\subsection{Image generation}
  In the simulation for image generation,
  we used generative adversarial networks (GANs)
  ~\cite{goodfellow2014generative, radford2015unsupervised}.
  Basic GANs consists of two networks called ``Generator'' and ``Discriminator,''
  respectively.
  Generally,
  a generator includes transposed convolutional layers
  and a discriminator includes strided convolutional layers.
  For this reason,
  checkerboard artifacts are generated in both the forward- and backward-propagation
  of GANs.
  To avoid the artifacts, we applied the proposed layers to GANs,
  where filter kernels of the proposed layer were
  $\mytensor{K}^{(0)}, \mytensor{K}^{(1)}$, and $\mytensor{K}^{(2)}$.

  Tables \ref{tab:generator} and \ref{tab:discriminator} show
  the architectures of a generator and a discriminator used in the simulation,
  respectively,
  where Leaky ReLU~\cite{xu2015empirical} was utilized as activation functions
  in these networks.
  \begin{table}[!t]
    \centering
    \footnotesize
    \caption{Generator architecture.
      \textit{ch} denotes the number of output channels.}
    \begin{tabular}{cccc} \hline\hline
      Layer & Stride $s$ & Kernel size & \textit{ch} \\ \hline
      Trans. conv. + BN + Leaky ReLU & 2 & $4 \times 4$ & 512 \\
      Trans. conv. + BN + Leaky ReLU & 2 & $4 \times 4$ & 512 \\
      Trans. conv. + BN + Leaky ReLU & 2 & $4 \times 4$ & 512 \\
      Trans. conv. + BN + Leaky ReLU & 2 & $4 \times 4$ & 256 \\
      Trans. conv. + BN + Leaky ReLU & 2 & $4 \times 4$ & 128 \\
      Trans. conv. + BN + Leaky ReLU & 2 & $4 \times 4$ & 64 \\
      Conv. + Tanh & 1 & $3 \times 3$ & 3 \\
      \hline
    \end{tabular}
    \label{tab:generator}
  \end{table}
  \begin{table}[!t]
    \centering
    \footnotesize
    \caption{Discriminator architecture.
      \textit{ch} denotes the number of output channels.}
    \begin{tabular}{cccc} \hline\hline
      Layer & Stride $s$ & Kernel size & \textit{ch} \\ \hline
      Conv. + BN + Leaky ReLU & 2 & $4 \times 4$ & 8 \\
      Conv. + BN + Leaky ReLU & 2 & $4 \times 4$ & 32 \\
      Conv. + BN + Leaky ReLU & 2 & $4 \times 4$ & 64 \\
      Conv. + BN + Leaky ReLU & 2 & $4 \times 4$ & 64 \\
      Conv. & 1 & $4 \times 4$ & 1 \\
      \hline
    \end{tabular}
    \label{tab:discriminator}
  \end{table}
  We trained GANs with 10 epochs by using 202,599 images in the CelebA~\cite{liu2015deep} dataset
  in order to generate face images from 100-dimensional random vectors that follow
  the standard normal distribution,
  where the images were resized to $64 \times 64$ pixels.
  The Adam optimizer \cite{kingma2014adam}
  with an initial learning rate of 0.08, $\beta_1=0.5$, and $\beta_2=0.9$
  was utilized for training both the generator and the discriminator.

\begin{figure}[!t]
  \centering
  \begin{subfigure}[t]{0.24\hsize}
    \centering
    \includegraphics[width=\columnwidth]{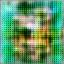}
    \caption{without fixed layers \label{fig:conv_gan}}
  \end{subfigure}
  \begin{subfigure}[t]{0.24\hsize}
    \centering
    \includegraphics[width=\columnwidth]{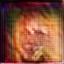}
    \caption{Proposed ($\mytensor{K}^{(0)}$) \label{fig:prop_gan_0}}
  \end{subfigure}
  \begin{subfigure}[t]{0.24\hsize}
    \centering
    \includegraphics[width=\columnwidth]{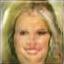}
    \caption{Proposed ($\mytensor{K}^{(1)}$) \label{fig:prop_gan_1}}
  \end{subfigure}
  \begin{subfigure}[t]{0.24\hsize}
    \centering
    \includegraphics[width=\columnwidth]{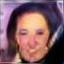}
    \caption{Proposed ($\mytensor{K}^{(2)}$) \label{fig:prop_gan_2}}
  \end{subfigure}
  \vspace{-2ex}
  \caption{Generated images by using GAN.
    Filter kernel $\mytensor{K}^{(d)}$ of proposed fixed layer
    is controlled by order of smoothness $d$.
    \label{fig:generation}}
\end{figure}

  Figure \ref{fig:generation} illustrates an example of face images generated
  by using trained GANs.
  From the figure, checkerboard artifacts in generated images were suppressed
  by using the proposed layer
  although conventional GANs without the proposed layer caused the artifacts.
  Furthermore, the use of the proposed layers enables us to
  improve the quality of generated images.
  By comparing
  figures \ref{fig:generation}(\subref{fig:prop_gan_0})，(\subref{fig:prop_gan_1})，
  and (\subref{fig:prop_gan_1}),
  it is confirmed that a sharp and clear image was generated when
  filter kernels $\mytensor{K}^{(1)}$ and $\mytensor{K}^{(2)}$ were used.
  From these results,
  the fixed convolutional layer having a smoother filter kernel
  provides GANs with a better performance.

\section{Conclusion}
  We proposed a novel fixed convolutional layer with an order of smoothness.
  The proposed layer can not only perfectly avoid
  checkerboard artifacts caused by transposed convolutional layers
  and strided convolutional layers,
  but also enhance the performance of CNNs including GANs.
  A filter kernel in the proposed layer
  is calculated by using a zero-order hold kernel
  and a parameter so-called \textit{the order of smoothness}.
  From an image-classification experiment,
  it was confirmed that avoiding checkerboard artifacts by the proposed layer
  improved the classification performance of all four CNNs
  that we used in the experiment.
  In addition,
  an image-generation experiment demonstrated that
  a smoother fixed convolutional layer enables us to improve
  the quality of generated images by GANs while avoiding
  checkerboard artifacts,
  for the first time.
  In future work,
  we will further evaluate the proposed layer by more detailed simulations
  using other network architectures such as U-Net
  and discuss its effectiveness theoretically.

%

\end{document}